\documentclass[journal]{IEEEtran}
\IEEEoverridecommandlockouts
\usepackage{algorithmic}
\usepackage{array}
\usepackage{cite}
\usepackage{amsmath,amssymb,amsfonts}
\usepackage{algorithmic}
\usepackage{graphicx}
\usepackage{textcomp}
\usepackage{caption}
\usepackage{subcaption}
\usepackage[dvipsnames]{xcolor}

\def\BibTeX{{\rm B\kern-.05em{\sc i\kern-.025em b}\kern-.08em
    T\kern-.1667em\lower.7ex\hbox{E}\kern-.125emX}}
    
\def\pp[#1]{\left( #1 \right)}

\begin{document}
%
\title{Wideband Time Frequency Coding}
%
%
%

\author{Kathleen Yang,~\IEEEmembership{Student member, IEEE,}
        Salman Salamatian,~\IEEEmembership{}
        Rafael G. L. D'Oliveira,~\IEEEmembership{Member, IEEE,}
        Muriel M\'{e}dard,~\IEEEmembership{Fellow,~IEEE}
\thanks{
Kathleen Yang and Muriel Médard are with
the Research Laboratory of Electronics, MIT, Cambridge, MA 02139 USA
(e-mail: klyang@mit.edu; medard@mit.edu).

Salman Salamatian was with MIT, Cambridge, MA 02139 USA. He
is now with D. E. Shaw Group, New York, NY 10036 USA (e-mail:
salman.salamatian@gmail.com)

Rafael G. L. D’Oliveira was with MIT, Cambridge, MA 02139 USA. (e-mail: rafaeld@mit.edu)}
\thanks{This work was supported by the Office of Naval Research under the Battelle prime contract N6833518C0179.}
\thanks{Manuscript received ; revised .}
\thanks{Part of this work has been previously presented in part in PIMRC 2019 \cite{wtfc}.
In this work, we expand into further detail on how changing the parameters of WTFC impacts its capacity, and introduce large-scale fading and the shadowing effect to the channel model.}}

\maketitle

\begin{abstract}
In the wideband regime, the performance of many of the popular modulation schemes such as code division multiple access and orthogonal frequency division multiplexing falls quickly without channel state information.
Obtaining the amount of channel information required for these techniques to work is costly and difficult, which suggests the need for schemes which can perform well without channel state information.
In this work, we present one such scheme, called wideband time frequency coding, which achieves rates on the order of the additive white Gaussian noise capacity without requiring any channel state information.
Wideband time frequency coding combines impulsive frequency shift keying with pulse position modulation, which allows for information to be encoded in both the transmitted frequency and the transmission time period. 
On the detection side, we propose a non-coherent decoder based on a square-law detector, akin to the optimal decoder for frequency shift keying based signals.
The impacts of various parameters on the symbol error probability and capacity of wideband time frequency coding are investigated, and the results show that it is robust to shadowing and highly fading channels.
When compared to other modulation schemes such as code division multiple access, orthogonal frequency division multiplexing, pulse position modulation, and impulsive frequency shift keying without channel state information, wideband time frequency coding achieves higher rates in the wideband regime, and performs comparably in smaller bandwidths.
\end{abstract}


%
\IEEEpeerreviewmaketitle

\section{Introduction}
%
%
%
%
\IEEEPARstart{T}{he} desire for higher data rates and the currently crowded frequency spectrum has led to an interest in the wideband regime, where there is more bandwidth available \cite{mmwave_rappaport}. 
There are various challenges associated with the transition to the wideband regime, which can detrimentally impact previously used signaling schemes such as code division multiple access (CDMA) and orthogonal frequency division multiplexing (OFDM).
One such challenge is obtaining accurate and reliable channel state information, which CDMA and OFDM require in order to achieve good data rates \cite{lte_ofdma, lte_3gpp}.
Doing so is difficult due to the fading and noise associated with the wideband regime \cite{6G_non_coherent}.
Without channel state information, the capacities of CDMA and OFDM fall drastically. 
This naturally leads to a renewed interest in investigating signaling schemes that do not require any CSI for usage in the wideband regime.

Theoretical results such as \cite{gomez_cuba} demonstrate that the capacity of a signaling scheme where the signal is not concentrated or peaky will inevitably fall to zero as a function of the bandwidth once a critical bandwidth is passed. For example, the capacity of CDMA and other direct sequence spread spectrum signaling (DSSS) schemes approach zero as the bandwidth increases due to fourth moment constraints \cite{cdma_wideband, fourthegy}, and similarly, the capacity of OFDM approaches zero as the bandwidth increases past a critical bandwidth \cite{OFDM_cap}.
On the other hand, a signal concentrated in both time and frequency performs well in doubly dispersive channels, which are associated with the wideband regime \cite{kennedy_fading}.
One strategy to generate these concentrated or impulsive signals, is by incorporating a duty cycle to the signaling scheme \cite{gomez_cuba, porrat_wideband, ppm_wideband, porrat_dsss, teletar_pfsk,  single_two_tone_fsk}.
Indeed, the probability of error associated with any rate below the capacity of a fading dispersive channel can be made to be arbitrarily small through a careful choice of the duty cycle and inter-symbol time \cite{gallager_fading}.
Thus, there is motivation to explore impulsive signaling schemes for the wideband regime when there is no CSI.

The performance of various signaling schemes with incorporated duty cycles in the wideband regime without CSI has been previously investigated.
Pulse position modulation (PPM), which is a signaling scheme with an inherent duty cycle, was investigated for usage in the wideband regime.
It was shown that the capacity of PPM with repetition coding achieves rates close to the wideband capacity limit as the signal-to-noise ratio (SNR) increases, and once high SNR regimes are reached, the capacity of PPM levels off from limitations on the smallness of the symbol times due to the guard interval \cite{ppm_wideband}.
The performance of DSSS signaling schemes with a duty cycle were investigated in scenarios where the number of channel paths was large.
Unlike DSSS schemes without a duty cycle, the capacity of DSSS schemes with a duty cycle is more resilient to an increase in the number of channel paths \cite{porrat_dsss}, and therefore to the larger bandwidths.
OFDM with a duty cycle was investigated in multiple input multiple output (MIMO) scenarios in the wideband regime, and rates close to the capacity bound are achievable with critical duty cycles, at which the capacity of the scheme peaks for the given parameters \cite{gomez_cuba}.
Frequency shift keying (FSK) with a duty cycle, which is otherwise known as impulsive FSK (I-FSK), approaches the capacity limit as the bandwidth goes to infinity and the duty cycle approaches zero \cite{teletar_pfsk}.
In the finite bandwidth and non-zero duty cycle scenario, single and two tone I-FSK can achieve within 2 dB of the energy-limited capacity bound with an optimal duty cycle \cite{single_two_tone_fsk}.

In much of the prior research on the performance of signaling schemes with a duty cycle in the wideband regime, the time duration in which the signal is transmitted is known to both the transmitter and the receiver.
However, information can be encoded in the time duration the signal was transmitted, as in the case of PPM.
In this work, we present a wideband time frequency coding (WTFC) modulation that combines single tone I-FSK and PPM.
In this modulation scheme, information can be encoded in both the duration of a signal transmission and the frequency of the transmitted signal.
A frequency tone is transmitted once per duty cycle with an increased amplitude, and the time period in which the signal is transmitted is unknown to the receiver.

There has been prior work on a combined FSK and PPM modulation, but primarily for applications outside the wideband regime. 
In optical communications, low-dimensional FSK and PPM modulations were used to reduce the required SNR compared to other modulations for similar bit error rates (BER) \cite{visible_light} and to reduce the power penalties for a given BER \cite{optical_label}.
A FSK and PPM modulation was used in low energy critical infrastructure monitoring networks in order to reduce the power requirement compared to an FSK modulation for equivalent BERs \cite{lecim}.

In comparison to these works, WTFC focuses on a high dimensional combined FSK and PPM modulation for usage in the wideband regime without CSI.
We demonstrate that WTFC performs well over a large range of bandwidths, is robust under various channel conditions, and can achieve rates on the order of the capacity bound.
The capacity of WTFC remains stable as the bandwidth becomes large and is greater than the capacity of other modulations such as CDMA with a duty cycle, OFDM with a duty cycle, and PPM, whose capacities go to zero at larger bandwidths.
When comparing WTFC to I-FSK, we show that WTFC outperforms I-FSK when the two modulations have the same duty cycle.

This paper is organized as follows: 
In Section \ref{sec:signal_model}, we discuss the Rayleigh fading channel model with large-scale fading, and introduce the WTFC signal.
In Section \ref{sec:demod+channel_model}, we discuss the non-coherent square law receiver, and the discrete memoryless channel model used to find the capacity of WTFC.
In Section \ref{sec:results}, we first investigate how different parameters impact the probability of error and capacity of WTFC, and then compare the performance of WTFC to other modulations such as CDMA, OFDM, PPM, and I-FSK.
In Section \ref{sec:conclusion}, we give our concluding remarks.

The notations that are commonly used throughout this paper are listed in Table \ref{table:defs}.
\section{Signal Model}

\begin{table}[t!]
\centering
\normalsize
\begin{tabular}{|c |c |} 
 \hline
 Notation & Description \\ [0.5ex] 
 \hline\hline
 $N_0$ & Noise spectral density\\ \hline
 $\alpha$ & Small-scale fading factor \\ \hline
 $m$ & Large-scale fading factor \\ \hline
 $\sigma$ & Shadowing standard deviation\\ \hline
 $\Delta_f$ & Separation between adjacent frequencies  \\ \hline
 $M$ & Total number of frequencies available  \\\hline
 $B$ & Bandwidth \\\hline
 $T_s$ & Symbol time \\\hline
 $T_d$ & Delay spread \\ \hline
 $B_d$ & Doppler spread \\ \hline
 $\theta$ & Duty cycle \\ \hline
 $P_t$ & Transmit power \\ \hline
 $P_r$ & Receive power \\ \hline
\end{tabular}
\caption{Notations that are commonly used throughout this paper.}
\label{table:defs}
\end{table}

\label{sec:signal_model}

\subsection{Channel Model}
\label{sec:channel_model}
We consider a multipath fading channel model with and without log-normal path loss and shadowing.
For an input signal $x(t)$, the channel output $y(t)$ is \cite{proakis, sklar_textbook}
\begin{equation}
    y(t) = m(t) \sum_{r = 1}^{L}a_r(t)x(t-d_r(t)) + z(t)
    \label{eqn:unsimplified_channel}
\end{equation}
where $m(t)$ is the large-scale fading that includes both path loss and shadowing, $a_r(t)$ and $d_r(t)$ are the channel gain and delay associated with path $r$, respectively, L is the total number of paths, and $z(t)$ is the complex additive Gaussian white noise with noise spectral density $N_0$.

The large-scale fading term $m(t)$ can be represented as a constant $m$ that is time-independent \cite{sklar_textbook, rappaport_textbook}
\begin{equation}
\begin{split}
    m(t) = m = \sqrt{10^{-L_p(d)/10}}
\end{split}
\end{equation}
\begin{equation}
\begin{split}
    L_p(d) = 20\log_{10}\pp[\frac{4\pi d_0}{\lambda}] + 10\gamma\log_{10}\pp[\frac{d}{d_0}] + X_\sigma
    \label{eqn:path_loss}
\end{split}
\end{equation}
where $d$ is the distance between the transmitter and receiver, $d_0$ is the reference distance, $\lambda$ is the wavelength of the signal, $\gamma$ is the path loss coefficient, and $X_\sigma$ is a zero-mean Gaussian random variable with standard deviation $\sigma$.
In this work, we assume free-space path loss, but an experimentally derived path loss relationship may be used instead of the first term of (\ref{eqn:path_loss}) \cite{sklar_textbook}.

The channel is assumed to have block fading in both frequency and time.
With this assumption, the channel gain $a_r(t)$ and channel delay $d_r(t)$ remain constant throughout a coherence time $T_c$ and are independent from one coherence time period to the next \cite{proakis}.
The channel gain and delay are highly correlated throughout a coherence bandwidth $B_c$, and fade simultaneously within this coherence bandwidth.
Thus, we can express the channel gains and delays as constants where $a_r(t) = a_r$ and $d_r(t) = d_r$.
The delay spread $T_d$ is approximately the inverse of the coherence bandwidth, and the Doppler spread $B_d$ is approximately the inverse of the coherence time \cite{goldsmith_book}.
The delay spread is assumed to be constant, i.e., $\max d_r \ll T_d$.
For an underspread channel where $T_d \ll T_c$, we can rewrite (\ref{eqn:unsimplified_channel}) as \cite{proakis}
\begin{equation}
    y(t) = m\sum_{r=1}^{L}a_rx(t-d_r) + z(t)
    \label{eqn:multipath}
\end{equation}

\subsection{Wideband Time Frequency Coding Signal}
\label{sec:wtfc_model}

\begin{figure}[t]
    \centering
    \includegraphics[width = \linewidth]{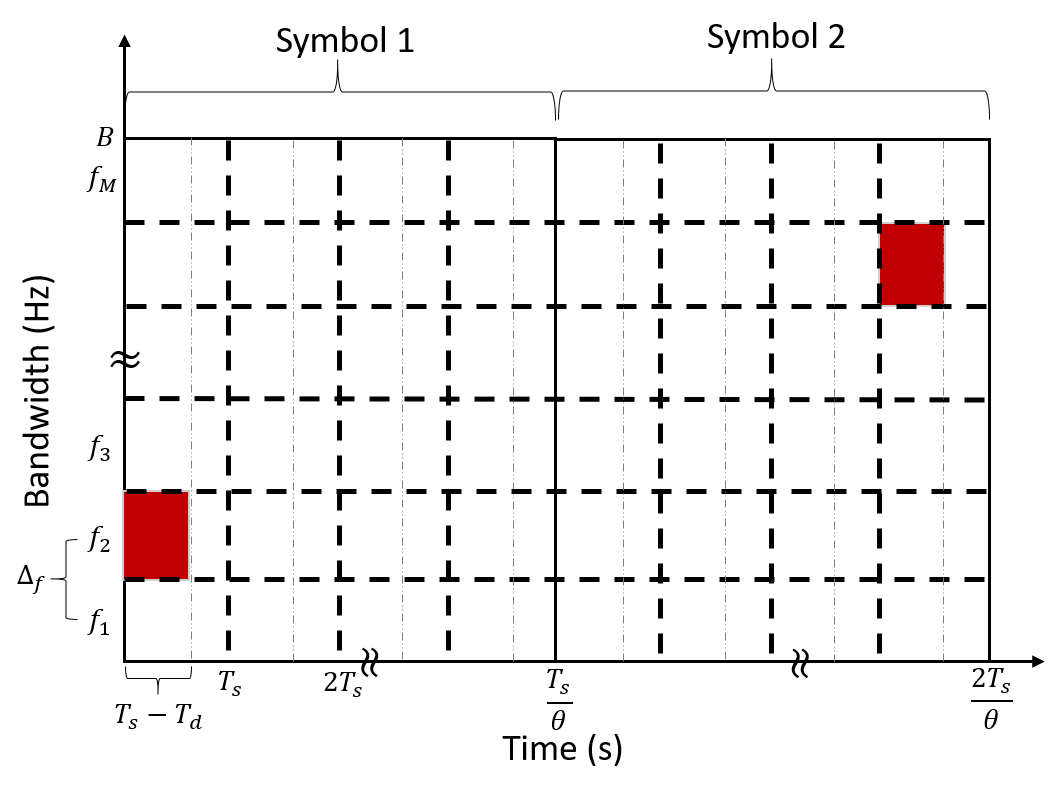}
    \caption{Grid depicting the transmission of two WTFC symbols with a duty cycle of $\theta$ and $M$ possible frequencies. Shaded regions denote the time period the symbols were transmitted in, as well as the transmitted frequencies. }
    \label{fig:wtfc_grid}
\end{figure}

WTFC is a modulation scheme that is inspired by I-FSK and combines FSK with PPM \cite{wtfc}.
The signal is comprised of a single frequency tone whose amplitude is increased by the square root of the inverse of the duty cycle, and is transmitted only once per duty cycle during any time period.
Information is encoded in both the frequency of the signal and the time duration that the signal is transmitted in.
Fig. \ref{fig:wtfc_grid} illustrates the transmission of two WTFC symbols with a duty cycle of $\theta$ and $M$ possible frequencies. 
It can be seen that the transmission of the WTFC signal is not fixed to a time period, and only a single frequency is transmitted once per duty cycle.

Given a bandwidth $B$, a symbol time $T_s$, and an adjacent frequency separation $\Delta_f = f_{i+1} - f_{i} = q/(T_s-T_d)$ where $q \in \mathbb{N}$, there are a total of $M = B(T_s-T_d)/q$ possible frequencies that can be transmitted.
The frequency separation $\Delta_f$ is a multiple of $1/(T_s-T_d)$ in order to maintain orthogonality and prevent spectral leakage, and the $q$ factor in $\Delta_f$ is chosen such that $\Delta_f \geq B_d$ in order to account for the Doppler spread when in a moving system.
The signal is transmitted for a duration of $T_s-T_d$ in order to account for the delay spread $T_d$ and to prevent the potential overlap of two time-adjacent signals at the receiver.
The delay spread $T_d$ is used as a guard time, and other guard times may be used as long as they are greater than the delay spread.
The transmission of the signal occurs once during the cycle time period of $[(p-1)T_s/\theta,\, pT_s/\theta)$ where $p\in \mathbb{N}$.

For simplicity, we consider the signal transmission over a cycle time of $[0, T_s/\theta)$.
The transmitted signal is \cite{wtfc}
\begin{equation}
    x_{l,k}(t) = 
    \begin{cases}
    A\mathrm{exp}(j2\pi f_lt),& kT_s\leq t \leq (k+1)T_s-T_d, \\
    0 ,& \mathrm{otherwise}.
    \end{cases} 
    \label{eqn:wtfc_transmission}
\end{equation}
where 
\begin{equation}
    A = \sqrt{\frac{P_tT_s}{\theta(T_s-T_d)}}
    \label{eqn:amplitude} \, ,
\end{equation}
is the amplitude of the signal, $f_l$ is the frequency that is being transmitted, $k$ is the index of the time period in which the signal is being transmitted where $k\in [0, 1/\theta -1]$, $\theta$ is the duty cycle, $T_s$ is the symbol time, and $P_t$ is the average transmit power.

\subsection{Channel Output}
\begin{figure}[t]
    \centering
    \includegraphics[width = \linewidth]{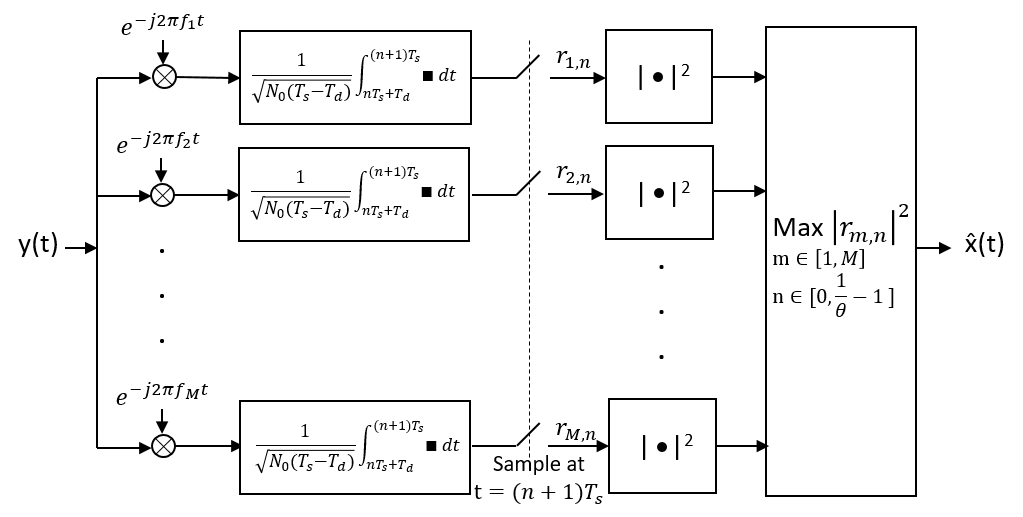}
    \caption{Square-law detector used for non-coherent detection of WTFC signals.}
    \label{fig:matched_filters}
\end{figure}

The multipath fading channel output can be found using the Central Limit Theorem when the number of paths, $L$, is large and the fading coefficients, $a_r$, are independent \cite{proakis}.
Given the channel described in (\ref{eqn:multipath}) and the transmitted signal $x_{l,k}(t)$, the channel output is \cite{wtfc}
\begin{equation}
    \resizebox{.44\textwidth}{!}{$\displaystyle y(t) = \begin{cases} m \alpha  A\mathrm{exp}(j2\pi f_lt) + z(t),& kT_s\leq t \leq (k+1)T_s, \\
    z(t),& \mathrm{otherwise.}
    \end{cases} $}
    \label{eqn:wtfc_channel_output}
\end{equation}
where $m$ is the large-scale fading factor, $\alpha$ is a zero-mean complex Gaussian random variable with $\mathbb{E}[|\alpha|^2] = 1$, $z(t)$ is the complex additive white Gaussian noise (AWGN) with power spectral density $N_0/2$.

It should be noted that the average received power of the signal $P_r$ can be related to the average transmit power $P_t$ through
\begin{equation}
\log(P_t) = \pp[2\log\pp[\frac{4\pi d_0}{\lambda}]+\gamma\log\pp[\frac{d}{d_0}]]+\log(P_r)
    \label{eqn:transmit_receive_power}
\end{equation}
Throughout this work we will be primarily using the averaged received power of the signal.
The relationship between the transmit and receive power will remain implicit.

\section{Demodulation and performance}
\label{sec:demod+channel_model}

\subsection{Demodulation}
\label{sec:demodulation}
A non-coherent receiver is used in order to avoid the costs associated with obtaining CSI.
The optimal non-coherent FSK receiver is the square-law detector, which is a Maximum Likelihood Estimator for FSK-based signals \cite{proakis}.
The square-law detector, which is shown in Fig. \ref{fig:matched_filters}, is comprised of matched filters for all possible frequencies.
The matched filters are sampled every $T_s$ seconds. 
The maximum of all the squared magnitudes of the outputs of the matched filters over the entire duty cycle is used to determine the frequency of the signal and the time period in which the signal was transmitted.

The output of the $m$-th matched filter associated with frequency $f_m$, where $m\in[1, M]$, over time period $nT_s \leq t \leq (n+1)T_s$, where $n\in [0, 1/\theta - 1]$, is denoted by $r_{m,n}$.
The output $r_{m,n}$ can be formally calculated as
\begin{equation}
\begin{split}
    r_{m,n} &= \frac{1}{\sqrt{N_0(T_s-T_d)}}\int_{nT_s + T_d}^{(n+1)T_s}y(t)\exp\pp[-j2\pi f_m t] dt \\
    &= \begin{cases}
    m\alpha A \sqrt{\frac{P_t T_s}{\theta N_0}} + w_{m,n}, & \text{if} \,(m,n) = (l,k), \\
    w_{m,n}, &\text{otherwise}.
    \end{cases}
    \label{eqn:mf_output}
\end{split}
\end{equation}
where $w_{m,n}$ is a circularly symmetric Gaussian random variable with variance 1.
Decisions are then made using the maximum of all $|r_{m,n}|^2$.
An error occurs when $|r_{l,k}|^2 < |r_{m,n}|^2$ for $(m,n) \neq (l,k)$.
See App. \ref{app:no_shadowing} for the simplification of $|r_{m,n}|^2$ when large-scale fading is not considered.

\subsection{Discrete Memoryless Channel Model}
\label{sec:dmc_model}

\begin{figure}[t]
    \centering
    \includegraphics[width = \linewidth]{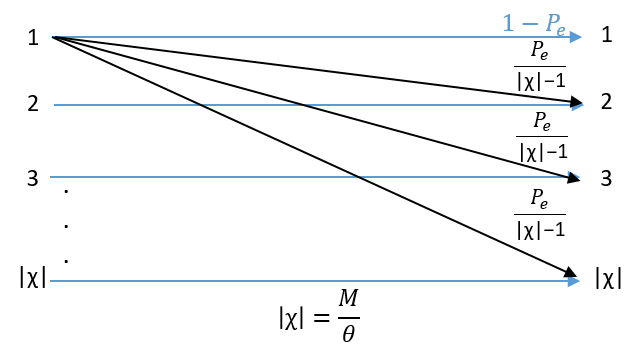}
    \caption{$M/\theta$-symmetric discrete memoryless channel model used for WTFC.}
    \label{fig:dmc}
\end{figure}

WTFC has been designed such that there is no temporal or spectral leakage, and all receiver outputs are independent.
The noise outputs also have identical distributions.
Therefore, we can model errors using the discrete memoryless channel model shown in Fig. \ref{fig:dmc}, which is a $M/\theta$ symmetric discrete memoryless channel.
The capacity of said channel is \cite{thomas_cover}
\begin{equation}
\begin{split}
     C =&\; \bigg(\mathrm{log_2}(M/\theta) + (1-P_e)\mathrm{log}_2(1-P_e)  \\&+ P_e\mathrm{log}_2\bigg(\frac{P_e}{M/\theta-1}\bigg)\bigg)\frac{\theta}{T_s}
    \label{eqn:channel_cap}
\end{split}
\end{equation}
where $P_e$ is the probability of an error which occurs when $|r_{l,k}|^2 < |r_{m,n}|^2$ at the receiver.
Note that there is an additional $\theta/T_s$ factor due to the duty cycle incorporated in WTFC - a symbol is transmitted only once every $T_s/\theta$ seconds.
In this work, the probability of an error is first found through simulation, and then used in (\ref{eqn:channel_cap}) to estimate the capacity bound.

\section{Numerical Results}
\label{sec:results}
In the simulations, we used $10^6$ iterations to find the probability of error and capacity associated with each signaling scheme (CDMA, PPM, OFDM, WTFC, and I-FSK) for the given parameters.
The inverse transform sampling method was used to find the probabilities of error and capacity associated with WTFC and I-FSK in order to reduce the simulation time.
See App. \ref{app:inverse_transform} for details on this method.
We use the AWGN capacity as a baseline comparison, as it is a valid upper bound on the capacity of multipath fading channels in both the finite and infinite bandwidth scenarios \cite{teletar_pfsk, rayleigh_cap}.

This section is split into two subsections.
We first explore how changing the parameters impacts the probability of error and capacity associated with WTFC.
We then compare the performance of WTFC to other modulation schemes under various channel conditions.

\subsection{Impact of changing parameters on WTFC}
\label{sec:changing_parameters}
There are various parameters that can impact the probability of error and capacity of WTFC.
In this subsection, we investigate how the duty cycle $\theta$, average receive power $P_r$, symbol time $T_s$, and the shadowing effect impact WTFC.

\begin{figure}[t]
    \centering
    \includegraphics[width = \linewidth]{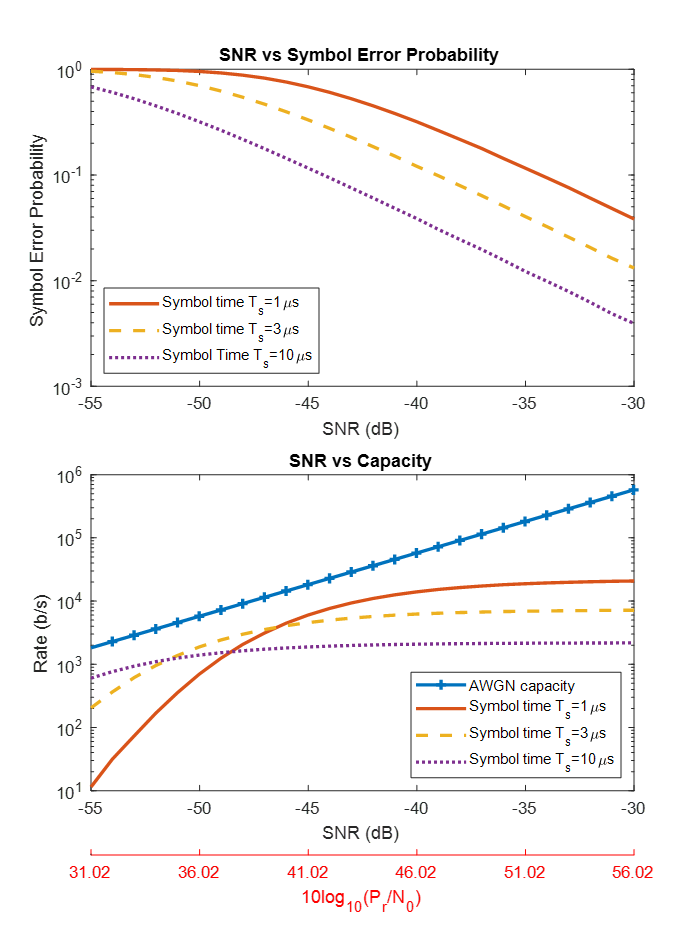}
    \caption{Impact of changing the average received power and the symbol time on the capacity of WTFC. The parameters used are:  duty cycle $\theta = 1/1000$, bandwidth $B = 400$ MHz, delay spread $T_d = 0.3\mu$s, and Doppler spread $B_d = 360$ Hz.}
    \label{fig:snr_plot}
\end{figure}

Fig. \ref{fig:snr_plot} shows how changing the averaged received power and symbol time affect the capacity of WTFC.
The x-axes are labeled with SNR, which can be calculated using $10\log_{10}(P_r/N_0B)$, and the ratio between the average received power and noise spectral density $10\log_{10}(P_r/N_0)$.
It can be seen that the symbol error probability decreases when the SNR increases.
This is due to the increased amount of energy that can be captured by the receiver when the SNR is larger.
The decrease in the symbol error rate when the SNR increases results in an increase in the capacity of WTFC.
However, a decreasing symbol error probability becomes less beneficial as the SNR grows large.
This is due to the innate limit on the total number of bits per symbol $(\lfloor\log_2(M/\theta)\rfloor)$ that can be transmitted.
Altering the symbol time can increase the capacity limit due to the shorter wait time between consecutive symbol transmissions, but the limitation on the capacity still exists regardless of the symbol time.

\begin{figure}[t]
    \centering
    \includegraphics[width = \linewidth]{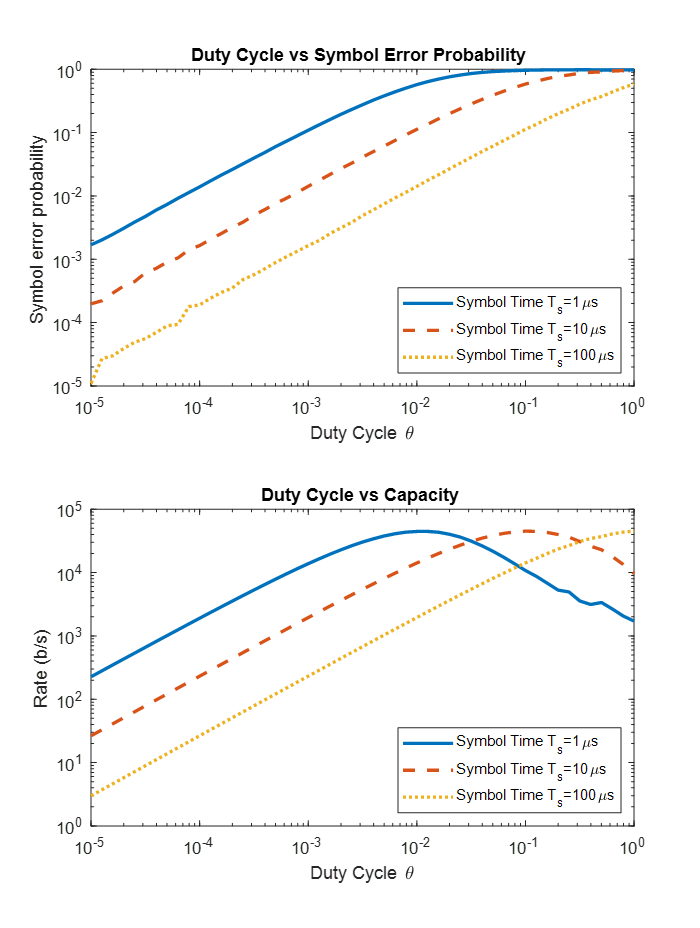}
    \caption{Impact of changing the duty cycle and the symbol time on the probability of error and capacity of WTFC. The parameters used are: average received power $P_r = 10^5$, noise spectral density $N_0 = 1$, bandwidth $B = 100$ MHz, delay spread $T_d = 0.3\mu$s, and Doppler spread $B_d = 360$ Hz.}
    \label{fig:duty_cycle_symbol_time}
\end{figure}

\begin{figure}[t]
    \centering
    \includegraphics[width = \linewidth]{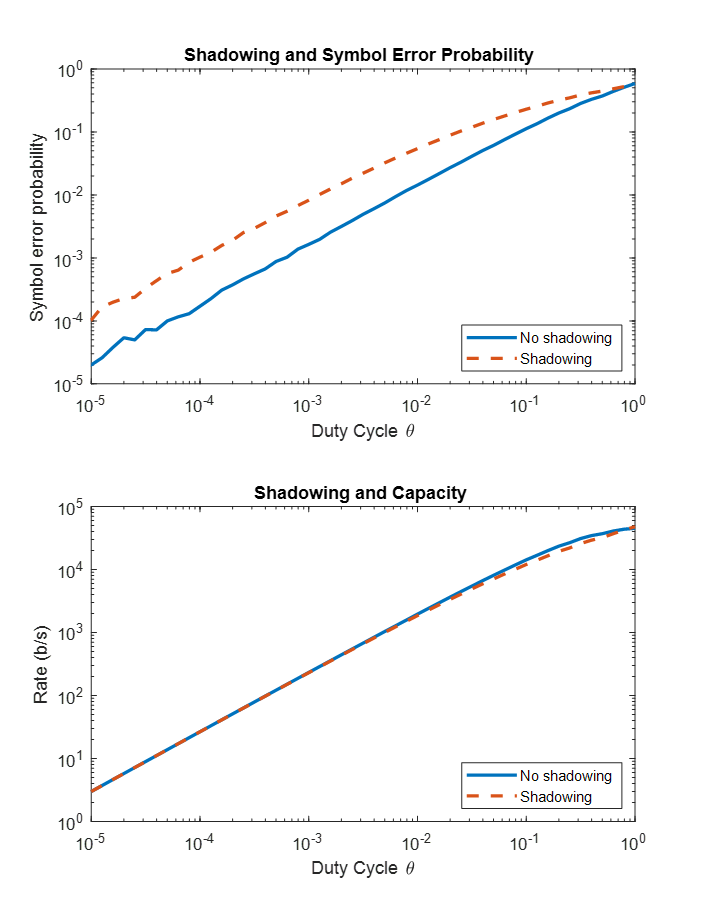}
    \caption{Impact of shadowing on the probability of error and capacity of WTFC. The parameters used are: average received power $P_r = 10^5$, noise spectral density $N_0 = 1$, bandwidth $B = 100$ MHz, symbol time $T_s = 100 \mu$s, delay spread $T_d = 0.3\mu$s, Doppler spread $B_d = 360$ Hz, and shadowing effect standard deviation $\sigma = 8$ dB.}
    \label{fig:shadowing_impact}
\end{figure}

It can be observed that the symbol error probability decreases as the symbol time increases.
This is due to the relationship in (\ref{eqn:mf_output}), where the output of the matched filters, $r_{m,n}$, is dependent on the square root of the symbol time.
For a given receive power, if the symbol time is increased, then the amount of energy that the square-law detector captures is also increased, which results in the smaller symbol error rate.
When the SNR is large and the symbol error probability is small, WTFC with smaller symbol times has a larger capacity limit.
The decreased wait time between consecutive symbol transmissions with smaller symbol times results in more transmitted bits per second.
This aspect is seen in (\ref{eqn:channel_cap}), where the capacity of the channel is inversely proportional to the symbol time. 
One interesting aspect to note is that the minimum ratio of the AWGN capacity to the WTFC capacity remains the same regardless of the symbol time. 
The symbol time should be carefully chosen for a target SNR and rate.

In addition to the symbol time, the duty cycle used also impacts the symbol error rate and the capacity.
Fig. \ref{fig:duty_cycle_symbol_time} illustrates how changing the symbol time and duty cycle affects the probability of error and capacity of WTFC.
As we have already discussed the impacts of the symbol time on the symbol error rate and capacity of WTFC, we focus on the effects of the duty cycle.
Decreasing the duty cycle results in a decrease in the symbol error probability due to the signal becoming more impulsive and having a larger amplitude, as can be seen in (\ref{eqn:amplitude}).
Despite the decrease in the duty cycle resulting in more possible time-frequency slots that a symbol can be represented by, the larger amplitude still results in a smaller symbol error probability.


There are two regimes of interest when looking into the impact of the duty cycle on the capacity of WTFC.
When the symbol error probability is greater than $1/2$, the capacity of WTFC increases when the duty cycle is decreased.
The reduction in the symbol error probability due to the amplitude of the signal being increased has a greater effect on the capacity than the increased wait time between consecutive symbols.
When the symbol error probability is less than $1/2$, the capacity of WTFC decreases with the decrease in the duty cycle.
In this regime, the increased wait time between consecutive symbols has a greater effect than the reduction in the symbol error probability.

One thing to note is that the maximum capacity that can be reached is the same for different pairs of duty cycle and symbol time.
This corresponds with what was stated in \cite{gomez_cuba}, where there exists a critical bandwidth occupancy that is dependent on symbol time, bandwidth, and duty cycle, at which the capacities of schemes are maximized.

\begin{figure}[t]
    \centering
    \includegraphics[width = \linewidth]{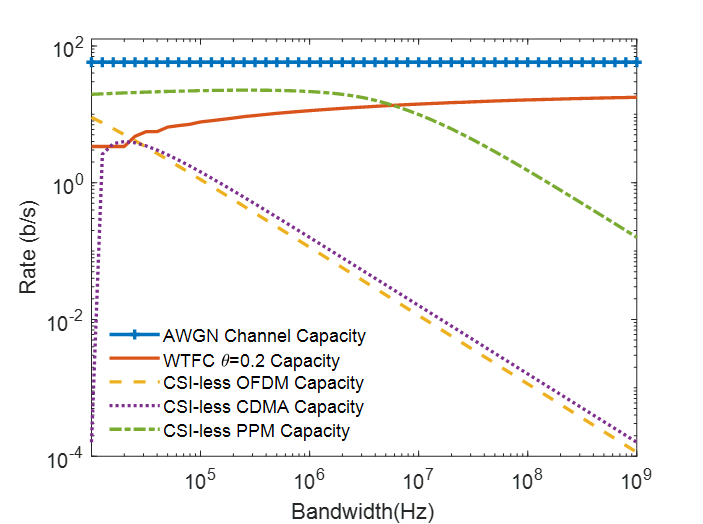}
    \caption{Comparison between WTFC, OFDM, CDMA, and PPM without CSI. The parameters used are: average received power $P_r = 40$, noise spectral density $N_0 = 1$, duty cycle $\theta = 1/5$, delay spread $T_d = 1\mu$s, and Doppler spread $B_d = 1000$ Hz.}
    \label{fig:other_scheme_comp}
\end{figure}


Fig. \ref{fig:shadowing_impact} shows the impact of shadowing on the symbol error probability and capacity of WTFC.
This figure was generated with the same parameters as Fig. \ref{fig:duty_cycle_symbol_time}, and with an additional shadowing effect with standard deviation $\sigma = 8$ dB and a symbol time of $T_s = 100 \mu$s.
The shadowing effect detrimentally impacts the symbol error probability, and its effects are most evident when the symbol error probability is small.
For example, at a duty cycle of $10^{-5}$, the the probability of a symbol error with shadowing is approximately 6 times greater than without shadowing, while at a duty cycle of one, the two symbol error probabilities are near equivalent.
Due to the increase of the symbol error probability from the shadowing effect, the capacity is lowered when there is shadowing.
The impact of the shadowing effect on the capacity is more evident when the symbol error probability is large.
WTFC is robust against shadowing, and shadowing has a small detrimental impact on its capacity.
When the symbol error probability is large, and near $p_e = 1/2$, the loss in capacity is approximately $3\%$, and when the symbol error probability is small, there is virtually no loss in capacity.

\subsection{Comparison between different modulations}

In this subsection, we investigate how WTFC performs in the wideband regime compared to other signaling schemes without CSI such as CDMA, OFDM, PPM, and I-FSK.
To find the capacity for CDMA without CSI, we use an upper bound on the wideband capacity, which can be found in Equation 94 from \cite{cdma_wideband}.
For OFDM without CSI, we use Equation 2 in \cite{OFDM_cap}.
For the capacity PPM with repetition coding, we use Equation 17 from \cite{ppm_wideband}.
We then compare the performance of WTFC versus I-FSK with different duty cycles to give insights into the benefits and downsides of altering the duty cycle.
For a final comparison, we compare the robustness of WTFC versus the other schemes in highly fading channels in order to show how WTFC is minimally impacted by channel conditions.

\begin{figure}[t]
    \centering
    \includegraphics[width = \linewidth]{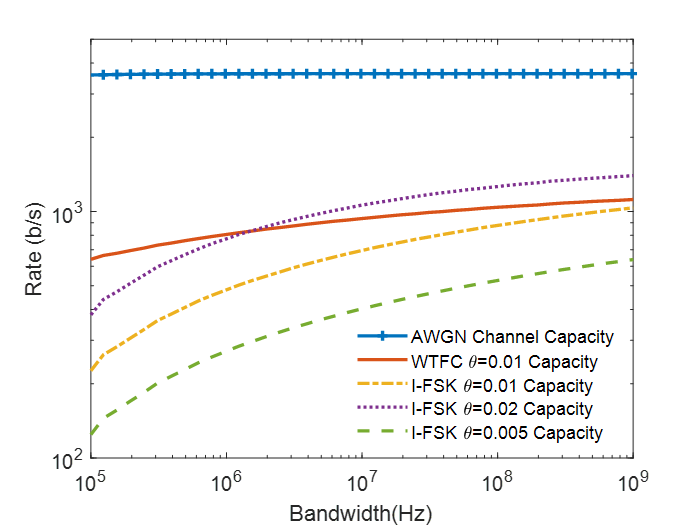}
    \caption{Comparison between WTFC with a duty cycle of $\theta = 1/100$ versus I-FSK with duty cycles of $\theta = 1/100, 1/50, 1/200$. The parameters used are: average received power $P_r = 10^{3.4}$, noise spectral density $N_0 = 1$, symbol time $T_s = 101\mu$s, delay spread $T_d = 20\mu$s, and Doppler spread $B_d = 360$ Hz.}
    \label{fig:wtfc_vs_ifsk}
\end{figure}

\begin{figure*}[t]
\centering
\begin{subfigure}{0.38\textwidth}
    \centering
    \includegraphics[width=\textwidth]{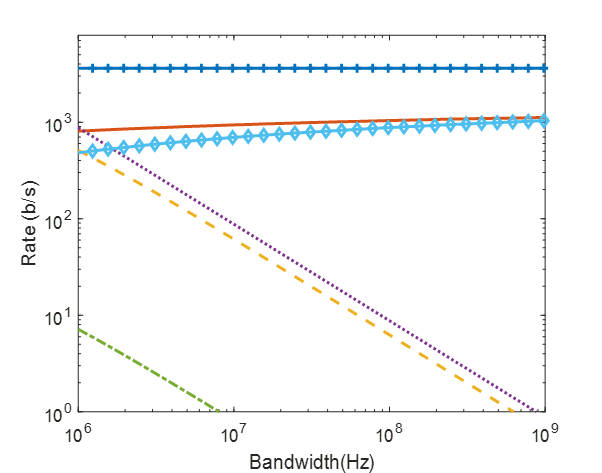}
    \caption{Performance of the signaling schemes with a Doppler spread of $B_d = 360$Hz, which corresponds to highway automobile speeds.}
    \label{fig:sf_highway_comp}
\end{subfigure}
\hfill
\begin{subfigure}{0.22\textwidth}
    \centering
    \includegraphics[width=\textwidth]{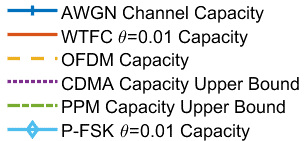}
    \label{fig:sf_comp_legend}
\end{subfigure}
\hfill
\begin{subfigure}{0.38\textwidth}
    \centering
    \includegraphics[width=\textwidth]{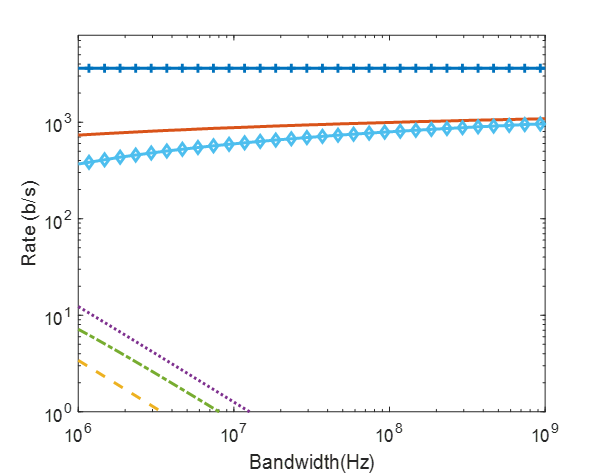}
    \caption{Performance of the signaling schemes with a Doppler spread of $B_d = 25 $kHz, which corresponds to aircraft speeds.}
    \label{fig:sf_airplane_comp}
\end{subfigure}

\caption{Comparison between the performance of WTFC, I-FSK, OFDM, CDMA, and PPM in different channel conditions. The parameters used are: average received power $P_r=10^{3.4}$, noise spectral density $N_0 = 1$, delay spread $T_d = 20 \mu $s, symbol time $T_s = 101 \mu $s, and duty cycle $\theta = 1/100$.}
\label{fig:fading_robustness}
\end{figure*}

Fig. \ref{fig:other_scheme_comp} shows a comparison between the capacities of WTFC, OFDM, CDMA, and PPM without CSI over bandwidths from 10 kHz to 1 GHz. 
Given the parameters used to find the capacity, CDMA outperforms WTFC when the bandwidth is small, but CDMA'a capacity begins to degrade when the bandwidth becomes greater than $\sim$30 kHz.
CDMA initially outperforms WTFC due to the small number of frequency slices at smaller bandwidths.
CDMA can take advantage of the correlation of the smaller number of frequency slices, but the correlation does not help improve its capacity once the number of frequency slices grows large \cite{cdma_wideband}.
This leads to CDMA's failure at larger bandwidths.
OFDM exhibits similar behavior to CDMA when it comes to comparing its performance to WTFC. 
It also outperforms WTFC until a bandwidth of $\sim$30 kHz, and past that bandwidth, WTFC outperforms it.
As the bandwidth grows, its capacity begins to approach zero.
This is due to the information being spread across time and frequency which results in a lack of an impulsive signal \cite{OFDM_cap}.
Without CSI and impulsive signals, CDMA and OFDM perform poorly and begin to fail as the bandwidth grows large.

In contrast to CDMA and OFDM, PPM does not perform as poorly.
PPM maintains a capacity close to the AWGN capacity bound until a bandwidth of $\sim$2 MHz is reached. 
Past this bandwidth, the capacity of PPM begins to gradually approach zero, similar to CDMA and OFDM's behavior.
This divergence from the capacity bound is due to restrictions on the symbol time given limitations on the guard time \cite{ppm_wideband}.

WTFC does not suffer the same gradual divergence away from the AWGN capacity bound that CDMA, OFDM, and PPM do.
The capacity of WTFC grows as the bandwidth increases.
This is due to the gradually increasing number of frequencies that can be transmitted when the bandwidth increases.
As the number of possible frequencies that can be transmitted increases, the number of bits per symbol also increases.
While the probability of a symbol error also increases with the bandwidth, the tradeoff between the increased symbol error probability and information gain still results in an overall increase in capacity.
This indicates that WTFC is robust in the wideband regime, unlike CDMA, OFDM, and PPM without CSI.
In addition, while WTFC may not perform as well in smaller bandwidths, its capacity is still of order of the other modulations' capacities, indicating that WTFC is versatile in multiple bandwidth regimes.

Fig. \ref{fig:wtfc_vs_ifsk} shows a comparison between the capacities of WTFC and I-FSK with various duty cycles in order to illustrate the impact of the duty cycle.
When comparing WTFC to I-FSK with the same duty cycle, WTFC outperforms I-FSK. 
This is due to the larger number of bits per WTFC symbol.
The extra information encoded in the time duration that the signal is transmitted in allows for WTFC to achieve a higher rate than I-FSK of the same duty cycle despite the increased symbol error probability.
When comparing WTFC to I-FSK with different duty cycles, it can be seen that I-FSK with a larger duty cycle has a larger capacity.
Increasing the duty cycle increases the symbol error probability, but reduces the inter-symbol time, which allows for an increase in the capacity.
However, even with an increase in I-FSK's duty cycle, WTFC still outperforms I-FSK in smaller bandwidths due to information being encoded in time.\footnote{It should be noted that this portion of this work was previously published in \cite{wtfc}, and a different trend was showed when comparing WTFC to I-FSK with different duty cycles. The mistake that resulted in this trend has been corrected in the journal version.}

Fig. \ref{fig:fading_robustness} compares the performance of WTFC versus CDMA, OFDM, PPM, and I-FSK in challenging channel conditions where the delay and Doppler spreads are large. 
Fig. \ref{fig:sf_highway_comp} compares the schemes under channel conditions that would be observed at highway automobile speeds in San Francisco, and Fig. \ref{fig:sf_airplane_comp} compares the schemes under channel conditions that would be observed at aircraft speeds in San Francisco.
We observe that the capacity of the non-impulsive schemes, CDMA and OFDM, are highly dependent on the channel conditions.
OFDM and CDMA perform better in channels that are less challenging, such as channels with a smaller Doppler spread.
In contrast, the impulsive schemes, WTFC, I-FSK, and PPM, are robust against the channel conditions.
While the capacities of these schemes also become smaller when the Doppler spread is larger, the decrease in the capacity is small compared to the decrease in capacity that CDMA and OFDM experience.

The decrease in the capacity of WTFC when the Doppler spread becomes large is primarily due to the required frequency separation $\Delta_f$.
Recall that the required frequency separation can be calculated from the symbol time $T_s$ and delay spread $T_d$ using $\Delta_f = q/(T_s-T_d)$ where $q \in \mathbb{N}$ with the additional condition that the frequency separation is greater than the Doppler spread $\Delta_f \geq B_d$ to prevent spectral leakage.
In Fig. \ref{fig:sf_highway_comp}, we have $\Delta_f \approx 12$ kHz, where $q=1$ satisfies the condition that $\Delta_f \geq B_d = 360$ Hz.
However, in Fig. \ref{fig:sf_airplane_comp}, the Doppler spread is $B_d = 25$ kHz, and we must choose $q = 3$ so that the condition $\Delta_f \geq B_d$ is satisfied.
This reduces the number of possible frequencies that can be transmitted, which results in a smaller number of bits per symbol, and a decrease in capacity.
Despite this decrease in capacity, it can still be seen that WTFC is robust under challenging channel conditions.

\section{Conclusion}
\label{sec:conclusion}
We presented a wideband time frequency coding scheme that combines I-FSK with PPM.
We investigated how the SNR, duty cycle, channel conditions, and shadowing effect impact its symbol error probability and capacity.
WTFC is robust against shadowing and challenging conditions, and can achieve rates of order of the AWGN capacity bound.
These aspects of WTFC show promise for the wideband regime.

We compared the performance of WTFC to CDMA, OFDM, PPM, and I-FSK over a large range of bandwidths and in different channel conditions.
WTFC performs well in large bandwidths, while CDMA, OFDM, and PPM begin to fail as the bandwidth grows large.
WTFC also performs well under challenging channels, while CDMA and OFDM fail in these channels.
In comparison to I-FSK, WTFC outperforms I-FSK when the two schemes have the same duty cycle.
When the duty cycles of I-FSK and WTFC differ, WTFC outperforms I-FSK at smaller bandwidths.

Future work would include performing over-the-air testing of WTFC. 
Proof of concept of I-FSK has already been demonstrated in \cite{milcom}, and shows the viability of an impulsive FSK-based signaling scheme under Federal Communications Commissions restrictions.
PPM has also been experimentally validated in optical and wireless communications, showing that PPM can be implemented \cite{ppm_optics, ppm_wireless}.
Given that WTFC is a combination of I-FSK and PPM, there is promise for over-the-air testing of WTFC to be shown in the future.

\section*{Acknowledgment}
We would like to thank Vedat and Assia Eyuboglu for their generous donation and support for this project.

\appendices
\allowdisplaybreaks
\section{Inverse transform sampling}
\label{app:inverse_transform}
In order to reduce the simulation time required to generate probabilities of error and capacity for WTFC, inverse transform sampling was used.
For simulations with $10^6$ iterations, the required number of random variable generations without inverse transform sampling would be $10^6 \times M/\theta$ where $M$ is the number of frequencies and $\theta$ is the duty cycle. 
With a small $\theta$ and a large $M$, the number of random variables that needs to be generated becomes large, and can take a large amount of computation time.
With inverse transform sampling and derivations of the distributions of the outputs of the receiver, the number of random variables that are generated can be reduced to the order of $10^6$.
Inverse transform sampling can be used both when there is and isn't large-scale fading.

\subsection{Simplification of receiver output without large-scale fading}
\label{app:no_shadowing}
While $|r_{m,n}|^2$ can be calculated directly from $r_{m,n}$, for the case where large-scale fading is not considered, it is faster to use the simplification of $|r_{m,n}|^2$ in simulation.

Without the large-scale fading term $m$, (\ref{eqn:mf_output}) is
\begin{equation}
\begin{split}
    r_{m,n} = \begin{cases}
    \alpha A \sqrt{\frac{P_t T_s}{\theta N_0}} + w_{m,n}, & \text{if} \,(m,n) = (l,k), \\
    w_{m,n}, &\text{otherwise}.
    \end{cases}
    \label{eqn:mf_output)_small_scale}
\end{split}
\end{equation}

Recall that $\alpha$ and $w_{m,n}$ are circularly symmetric complex Gaussian random variables with variance 1. 
The sum of two circularly symmetric complex Gaussian random variables remains a circularly symmetric complex Gaussian random variable, with a different variance.
When $(m,n) = (l,k)$, the variance of $r_{m,n}$ is $\frac{P_tT_s}{\theta N_0} + 1$, and when $(m,n) \neq (l,k)$ the variance of $r_{m,n}$ is $1$.

Multiple steps are used to simplify $|r_{m,n}|^2$ and find its distribution.
We first find the distribution of a square of a Gaussian random variable, and then find the distribution of the sum of two Gaussian random variables.
We generalize this process by using Gaussian random variables with variance $\sigma^2$.

\textbf{Step 1}: Square of Gaussian random variable $X=X_1^2$ where $X_1\sim\mathcal{N}(0, \sigma^2)$
\begin{equation}
\begin{split}
    F(x) & = P(X \leq x)\\
    & = P(X_1^2 \leq x) \\
    & = P(-\sqrt{x} \leq X_1 \leq \sqrt{x}) \\
    & = P(X_1 \leq \sqrt{x}) - P(X_1 \leq -\sqrt{x}) \\
    & = F_1(\sqrt{x}) - F_1(-\sqrt{x}) 
\end{split}
\end{equation}
\begin{equation}
\begin{split}
    f(x) & = \frac{dF(x)}{dx} \\
    & = \frac{1}{2\sqrt{x}}(F_1'(\sqrt{x}) + F_1'(-\sqrt{x})) \\
    & = \frac{1}{2\sqrt{x}}(f_1(\sqrt{x}) + f_1(-\sqrt{x})) \\
    & = \frac{1}{2\sqrt{x}}\frac{1}{\sqrt{2\pi\sigma^2}}2e^{-\frac{x}{2\sigma^2}} \\
    & = \frac{1}{\sqrt{2\pi\sigma^2x}}e^{-\frac{x}{2\sigma^2}}
\end{split}
\end{equation}

\textbf{Step 2}: Sum of two squared Gaussian random variables $Z=X_1^2+X_2^2$ where $X_1, X_2\stackrel{iid}{\sim}\mathcal{N}(0, \sigma^2)$
\begin{equation}
\begin{split}
    f(z) & = \int_{0}^{\infty}f_1(z-y)f_2(y)dy \\
    & = \int_{0}^{\infty}\frac{1}{\sqrt{2\pi\sigma^2}}(z-y)^{-\frac{1}{2}}e^{-\frac{z-y}{2\sigma^2}}\frac{1}{\sqrt{2\pi\sigma^2}}y^{-\frac{1}{2}}e^{-\frac{y}{2\sigma^2}}dy \\
    & = \frac{1}{2\pi\sigma^2}e^{-\frac{z}{2\sigma^2}}\int_{0}^{z}(z-y)^{-\frac{1}{2}}y^{-\frac{1}{2}}dy \\
    & = \frac{1}{2\pi\sigma^2}e^{-\frac{z}{2\sigma^2}}\int_{0}^{z}2(z-y)^{-\frac{1}{2}}d\sqrt{y} \\
    & = \frac{1}{2\pi\sigma^2}e^{-\frac{z}{2\sigma^2}}\int_{0}^{\sqrt{z}}2(z-x^2)^{-\frac{1}{2}}dx \\
    & = \frac{1}{2\pi\sigma^2}e^{-\frac{z}{2\sigma^2}}*\pi \\
    & = \frac{1}{2\sigma^2}e^{-\frac{z}{2\sigma^2}}
\end{split}
\end{equation}
\begin{equation}
\begin{split}
    F(z) & = \int_{0}^{z}f(z)dz \\
    & = \int_{0}^{z}\frac{1}{2\sigma^2}e^{-\frac{z}{2\sigma^2}}dz \\
    & = 1 - e^{-\frac{z}{2\sigma^2}}
\end{split}
\end{equation}
Note that the PDF and CDF in step 2 corresponds to an exponential random variable with $\lambda = 2\sigma^2$. That is, $Z \sim \text{Exp}(2\sigma^2)$.

With this, the squared outputs of the receiver are distributed as
\begin{equation}
    |r_{m,n}|^2 \sim \begin{cases}
    \text{Exp}(\mu), &\text{if} \, (m,n) = (l,k) \\
    \text{Exp}(1), &\text{otherwise}.
    \end{cases}
\end{equation}
where $\mu = \frac{P_tT_s}{\theta N_0} + 1$ is the mean of the exponential random variable when $(m,n) = (l,k)$.

\subsection{Distribution of the maximum of the noise}
\label{app:noise_max_dist}
While finding the distribution of $|r_{m,n}|^2$ reduces the number of random variable generations by approximately half, further reductions can be made.
In particular, we focus on finding the distribution of the maximum of the noise outputs.
The square-law rule still holds if we compare the non-noise receiver output versus the maximum of the noise receiver outputs instead of comparing the non-noise receiver output with all noise receiver outputs.
Using the maximum of all noise outputs, we only need to generate 1 random variable representing the noise for each iteration, compared to the previously required $M/\theta - 1$ random variables for each iteration.

We use the fact that the noise outputs are identically and independently distributed to derive the inverse CDF associated with the maximum of all noise outputs.
We first find the CDF associated with the maximum of N squared sums of Gaussian random variables, and then find the inverse CDF.

\textbf{Step 1}: CDF of the maximum of $N$ random variables that are the sum of two squared Gaussian random variables $X_i = X_{1i}^2 + X_{2i}^2$ where $X_{1i}, X_{2i} \stackrel{iid}{\sim}\mathcal{N}(0, \sigma^2) \forall i\in [1,N]$.
\begin{equation}
\begin{split}
    F(x) &= P(X\leq x) \\
    &= P(\max(X_1, X_2, ..., X_N) \leq x) \\ 
    &= \prod_{i=1}^{N}P(X_i \leq x) \\ 
    &= P(X_i \leq x)^N \\ 
    &= F_i(x)^N \\
    &= \left(1-\exp\left(-\frac{x}{2\sigma^2}\right)\right)^N
\end{split}
\end{equation}
where we use the identical distributions and independence of $X_i$ to simplify the CDF.

\textbf{Step 2}: Inverse CDF derivation
\begin{equation}
\begin{split}
    F(x) &= \left(1-\exp\left(-\frac{x}{2\sigma^2}\right)\right)^N \\
    F(x)^{\frac{1}{N}} &= \left(1-\exp\left(-\frac{x}{2\sigma^2}\right)\right) \\
    1 - F(x)^{\frac{1}{N}} &= \exp(-\frac{x}{2\sigma^2}) \\
    \ln\left(1-F(x)^{\frac{1}{N}}\right) &= -\frac{x}{2\sigma^2} \\
    x &= -2\sigma^2 \ln\left(1-F(x)^{\frac{1}{N}}\right) \\
    F^{-1}(p) &= -2\sigma^2 \ln\left(1-p^\frac{1}{N}\right)
\end{split}
\end{equation}
With the inverse CDF method, we can reduce the number of random variables that need to be generated to represent the receiver noise outputs from $N=M/\theta-1$ to 1 per iteration, which can greatly reduce the simulation time required.
\nocite{masters_thesis}



\bibliographystyle{IEEEtran}
\bibliography{main}

%








\end{document}